\documentclass[final]{svjour2}
\usepackage{graphicx}
\usepackage{rotating}
\usepackage{amssymb}
\usepackage{mathptmx}
\usepackage[numbers]{natbib}

\makeatletter
\journalname{Journal of Low Temperature Physics}

\bibpunct{}{}{,}{s}{}{,}

\begin{document}

\newcommand{\hdblarrow}{H\makebox[0.9ex][l]{$\downdownarrows$}-}
\title{Effect of the long-range Coulomb interaction on phase diagram of the
Kohn-Luttinger superconducting state in idealized graphene}

\author{M. Yu. Kagan$^{1,2}$ \and V. A. Mitskan$^{3,4}$ \and\\ M. M. Korovushkin$^{3}$}

\institute{$^1$P. L. Kapitza Institute for Physical Problems,
Russian Academy of Sciences,\\ Moscow 119334, Russia\\
$^2$National Research
University Higher School of Economics,\\
Moscow 109028, Russia\\
$^3$L. V. Kirensky Institute of Physics, Siberian Branch of
Russian Academy of Sciences,\\ Krasnoyarsk 660036, Russia\\
$^4$M. F. Reshetnev Siberian State Aerospace University,\\
Krasnoyarsk 660014, Russia\\
\email{kagan@kapitza.ras.ru}}

\date{29.06.2015}

\maketitle

\begin{abstract}

The effect of the long-range Coulomb interaction on the
realization of the Kohn-Luttinger superconductivity in idealized
monolayer doped graphene is studied. It is shown that the
allowance for the Kohn-Luttinger renormalizations up to the second
order in perturbation theory in the on-site Hubbard interaction
inclusively, as well as the intersite Coulomb interaction
significantly affects the competition between the superconducting
phases with the $f$-wave, $p+ip$-wave and $d + id$-wave symmetries
of the order parameter. It is shown that the account for the
Coulomb repulsion of electrons located at the next-nearest
neighboring atoms in such a system changes qualitatively the phase
diagram and enhances the critical superconducting temperature.


\end{abstract}

\section{Introduction}

At the present time, the possible development of superconductivity
in the framework of the Kohn-Luttinger mechanism~\cite{Kohn65},
suggesting the emergence of superconducting pairing in the systems
with the purely repulsive interaction~\cite{Kagan14b}, in graphene
under appropriate experimental conditions is widely discussed.
Despite the fact that intrinsic superconductivity so far has not
been observed in graphene, the stability of the Kohn-Luttinger
superconducting phase has been investigated and the symmetry of
the order parameter in the hexagonal lattice was identified. It
was found~\cite{Nandkishore12} that chiral
superconductivity~\cite{Black14} with the $d+id$-wave symmetry of
the order parameter prevails in a large domain near the Van Hove
singularity in the density of states (DOS). The competition
between the superconducting phases with different symmetry types
in the wide electron density range $1<n\leq n_{VH}$, where
$n_{VH}$ is the Van Hove filling, in graphene monolayer was
studied in papers~\cite{Kagan14a,Nandkishore14}. It was
demonstrated that at intermediate electron densities the Coulomb
interaction of electrons located on the nearest carbon atoms
facilitates implementation of superconductivity with the $f$-wave
symmetry of the order parameter, while at approaching the Van Hove
singularity, the superconducting $d+id$-wave pairing
evolves~\cite{Kagan14a,Nandkishore14}.

In this paper, we investigate the role of the Coulomb repulsion of
electrons located at the next-nearest neighboring carbon atoms in
the development of the Kohn-Luttinger superconductivity in an
idealized graphene monolayer disregarding the effect of the Van
der Waals potential of the substrate and both magnetic and
non-magnetic impurities. Using the Shubin-Vonsovsky (extended
Hubbard) model in the Born weak-coupling approximation, we
construct the phase diagram determining the boundaries of the
superconducting regions with different types of the symmetry of
the order parameter. It is shown that the account for the Coulomb
repulsion of electrons located at the next-nearest neighboring
sites of hexagonal lattice leads to a qualitative modification of
the phase diagram, as well as an increase in the critical
temperature of the transition to the superconducting state in the
system.

\section{Theoretical model}

In the hexagonal lattice of graphene, each unit cell contains two
carbon atoms. Therefore, the entire lattice can be divided into
two sublattices $A$ and $B$. In the Shubin-Vonsovsky
model~\cite{Shubin34}, the Hamiltonian for the graphene monolayer
taking into account the electron hoppings between the nearest
atoms, as well as the Coulomb repulsion between electrons located
at one, neighboring and next-nearest neighboring atoms in the
Wannier representation, has the form
\begin{eqnarray}\label{grapheneHamiltonian}
\hat{H}&=&\hat{H}_0+\hat{H}_{int},\\
\hat{H}_0&=&-\mu\Biggl(\sum_{f\sigma}\hat{n}^A_{f\sigma}+\sum_{g\sigma}
\hat{n}^B_{g\sigma}\Biggr)-t_1\sum_{
f\delta\sigma}(a^{\dag}_{f\sigma}b_{f+\delta,\sigma}+\textrm{h.c.}),\nonumber\\\label{monolayerH0}
\hat{H}_{int}&=&U\Biggl(\sum_f
\hat{n}^{A}_{f\uparrow}\hat{n}^{A}_{f\downarrow}+\sum_g
\hat{n}^{B}_{g\uparrow}\hat{n}^{B}_{g\downarrow}\Biggr)+V_1\sum_{
f\delta\sigma\sigma'}
\hat{n}^{A}_{f\sigma}\hat{n}^{B}_{f+\delta,\sigma'}\nonumber\label{monolayerHint}\\
&+&\frac{V_2}{2}\Biggl(\sum_{\langle\langle
fm\rangle\rangle\sigma\sigma'}\hat{n}^{A}_{f\sigma}\hat{n}^{A}_{m\sigma'}+\sum_{\langle\langle
gr\rangle\rangle\sigma\sigma'}\hat{n}^{B}_{g\sigma}\hat{n}^{B}_{r\sigma'}\Biggr).\nonumber
\end{eqnarray}
Here, operators $a^{\dag}_{f\sigma}(a_{f\sigma})$ create
(annihilate) an electron with spin projection $\sigma=\pm1/2$ at
the site $f$ of the sublattice $A$;
$\displaystyle\hat{n}^{A}_{f\sigma}=a^{\dag}_{f\sigma}a_{f\sigma}$
denotes the operator of the number of fermions at the $f$ site of
the sublattice $A$ (analogous notation is used for the sublattice
$B$). Vector $\delta$ connects the nearest atoms of the hexagonal
lattice. We assume that the position of the chemical potential
$\mu$ and the number of carriers $n$ in graphene monolayer can be
controlled by a gate electric field. In the Hamiltonian, $t_1$ is
the hopping integral between the neighboring atoms (hoppings
between different sublattices), $U$ is the parameter of the
Hubbard repulsion between electrons of the same atom with the
opposite spin projections, and $V_1$ and $V_2$ are the Coulomb
interactions between electrons of the neighboring and the
next-nearest neighboring carbon atoms in the monolayer. In the
Hamiltonian, the symbol $\langle\langle~\rangle\rangle$ means that
the summation is carried out only over the next-nearest neighbors.

\begin{figure}
\begin{center}
\includegraphics[%
  width=0.99\linewidth,
  keepaspectratio]{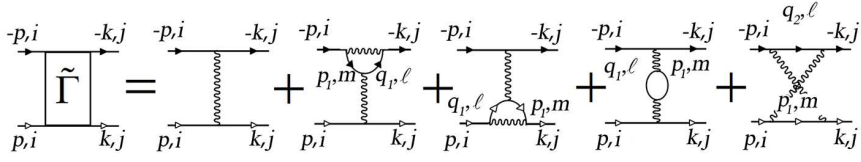}
\end{center}
\caption{The diagrams for the effective interaction of electrons
in graphene monolayer. Solid lines with light (dark) arrows
describe the Green functions for electrons with spin
$+{\textstyle{1 \over2}}$~($-{\textstyle{1 \over 2}}$) and
energies corresponding to the energy bands $E_{i}$ and $E_{j}$
($i$ and $j$ are equal to 1 or 2). Here momenta $\vec{q}_1 =
\vec{p}_1 + \vec{p} - \vec{k}$ and $\vec{q}_2 =
\vec{p}_1-\vec{p}-\vec{k}$ are introduced.} \label{diagrams}
\end{figure}

Proceeding to the momentum space and performing the Bogoliubov
transformation,
\begin{eqnarray}\label{uv}
\alpha _{i\vec{k}\sigma}= w_{i1}(\vec{k}){a_{ \vec{k}\sigma }} +
w_{i2}(\vec{k}){b_{\vec{k}\sigma }},\qquad i=1,2,
\end{eqnarray}
we diagonalize Hamiltonian $\hat{H}_0$, which acquires the form
\begin{eqnarray}
\hat H_0 =\sum\limits_{i=1}^2 \sum\limits_{ \vec{k}\sigma }
E_{i\vec{k}}
{\alpha_{i\vec{k}\sigma}^{\dag}\alpha_{i\vec{k}\sigma}}.
\end{eqnarray}
The two-band energy spectrum is described by the
expressions~\cite{Wallace47}
\begin{eqnarray}\label{spectra}
E_{1\vec{k}}=t_1|u_{\vec{k}}|-t_2f_{\vec{k}},\qquad
E_{2\vec{k}}=-t_1|u_{\vec{k}}|-t_2f_{\vec{k}},
\end{eqnarray}
where the following notation has been introduced:
\begin{eqnarray*}\label{f_k}
&&f_{\vec{k}}=2\cos(\sqrt{3}k_ya)+
4\cos\biggl(\frac{\sqrt{3}}{2}k_ya\biggr)\cos\biggl(\frac{3}{2}k_xa\biggr),\\
&&u_{\vec{k}}=\displaystyle\sum_{\delta}e^{i
\vec{k}\delta}=e^{-ik_xa}+
2e^{\frac{i}{2}k_xa}\cos\biggl(\frac{\sqrt{3}}{2}k_ya\biggr),\qquad
|u_{\vec{k}}|=\sqrt{3+f_{\vec{k}}}.
\end{eqnarray*}

The utilization of the Born weak-coupling approximation with the
hierarchy of model parameters
\begin{equation}\label{hierarchy}
W>U>V_1>V_2,
\end{equation}
where $W$ is the bandwidth in graphene monolayer (\ref{spectra}),
allows us to restrict the consideration only to the second-order
diagrams in the calculation of the effective interaction of the
electrons in the Cooper channel, and use the quantity
$\widetilde{\Gamma}(\vec{p}, \vec{k})$ for it. Note that this
quantity is determined by the diagrams presented in
Fig.~\ref{diagrams}. As far as the development of Cooper pairing
is determined by the properties of the energy spectrum and the
effective interactions of electrons in the vicinity of the Fermi
level~\cite{Gor'kov61}, we assume that the chemical potential of
the system falls into the upper energy band $E_{1\vec{k}}$ and
analyze the situation in which the initial and final momenta of
electrons in the Cooper channel also belong to this band. In this
paper, we perform the calculation of the superconducting phase
diagram in graphene following to the scheme we used in our
previous work~\cite{Kagan14a}.

\section{Results}

\begin{figure}
\begin{center}
\includegraphics[%
  width=1.03\linewidth,
  keepaspectratio]{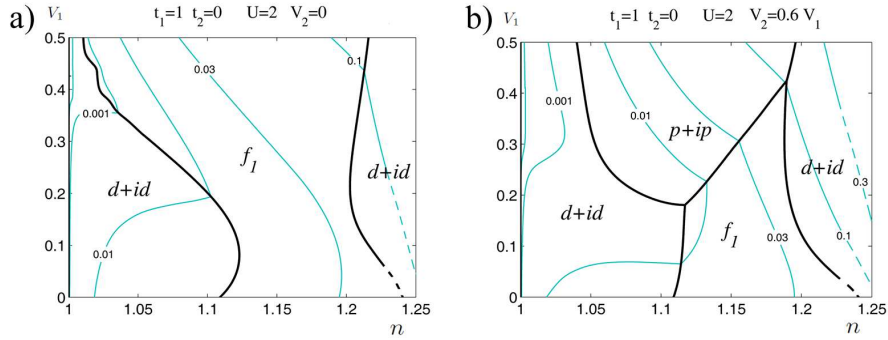}
\end{center}
\caption{(Color online) Superconducting phase diagram of an
idealized graphene monolayer at $U=2|t_1|$ for (a) $V_2=0$ and (b)
$V_2=0.6V_1$. Thin blue lines show the lines of constant
$|\lambda|$.} \label{PD}
\end{figure}

Figure~\ref{PD}a shows the calculated phase diagram of the
superconducting state in graphene monolayer as a function of the
carrier concentration $n$ and $V_1$ for the set of parameters
$U=2|t_1|$ and $V_2 = 0$. It can be seen that the phase diagram
consists of three regions. At low and high electron densities $n$,
the chiral superconducting $d+id$-wave pairing is
implemented~\cite{Black14,Nandkishore12}.
At the intermediate densities, the triplet $f$-wave pairing
occurs. With an increase of the intersite Coulomb interaction
$V_1$, at low electron densities, the superconducting $d+id$-wave
pairing is suppressed and the $f-$wave pairing is realized. In
Fig.~\ref{PD}a, thin blue lines corresponding to the lines of
constant absolute values of the effective coupling constant
$\lambda$, show that in the vicinity of $n_{VH}$ the values
$|\lambda|=0.1$.

In this paper, to avoid the consideration of the parquet
diagrams~\cite{Dzyaloshinskii88a,Dzyaloshinskii88b}, we analyze
the electron concentrations for the regions which are not too
close to the Van Hove singularity. In Fig.~\ref{PD}a, the dashed
lines show the boundaries between the different regions of the
superconducting pairing and the lines of $|\lambda|$ that are very
close to $n_{VH}$.
\begin{figure}
\begin{center}
\includegraphics[%
  width=0.55\linewidth,
  keepaspectratio]{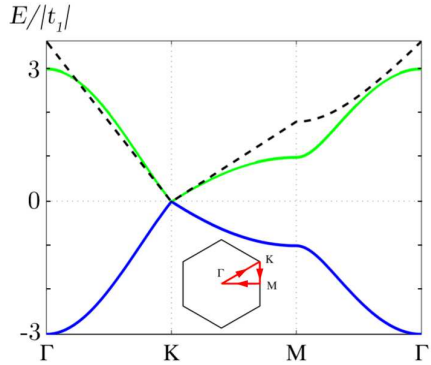}
\end{center}
\caption{(Color online) Energy spectrum of graphene monolayer
defined by (\ref{spectra}) (blue and green solid lines) and the
spectrum obtained in the framework of the Dirac approximation
(black dashed line). Subplot depicts the path around the Brillouin
zone.} \label{Dirac}
\end{figure}

Let us consider the influence of the Coulomb interaction $V_2$
between the electrons located at the next-nearest carbon atoms on
the phase diagram for graphene monolayer. In Fig.~\ref{PD}b
calculated for the fixed ratio between the parameters of the
intersite Coulomb interactions $V_2=0.6V_1$, one can see that an
account for $V_2$ leads to the qualitative modification of the
phase diagram. This modification involves the suppression of the
superconducting $f$-wave pairing at the large region of
intermediate electron densities and the realization of the chiral
superconducting $p+ip$-wave pairing. Additionally, when $V_2$ is
taken into account, the absolute values of the effective coupling
constant increases to $|\lambda|=0.3$. Consequently, it leads to a
significant increase in critical temperatures of the
superconducting transitions in idealized doped graphene. Note that
here we do not analyze the account for the electron hoppings to
the next-nearest carbon atoms $t_2$, because the consideration of
these hoppings for graphene monolayer does not significantly
modify the DOS in the carrier concentration regions between the
Dirac point and both points $n_{VH}$~\cite{Kagan14b,Kagan14a}.

Note that the Kohn-Luttinger superconducting pairing in graphene
never develops in the vicinity of the Dirac points. Our
calculations show that near these points, where the linear
approximation for the energy spectrum of graphene works pretty
well, the DOS is very low and the absolute values of the effective
coupling constant $|\lambda|<10^{-2}$. The higher values of
$|\lambda|$, which can indicate the development of the Cooper
instability at reasonable temperatures, arise at the electron
concentrations $n>1.15$. But at such concentrations, the energy
spectrum of the monolayer along the direction $KM$ of the
Brillouin zone (Fig.~\ref{Dirac}) significantly differs from the
Dirac approximation.

\section{Conclusions}

In conclusion, we have considered the conditions for the
superconducting pairing in the framework of the Kohn-Luttinger
mechanism in an idealized graphene monolayer, disregarding the
influence of the Van der Waals potential, as well as structural
disorder. The electronic structure of graphene is described in the
Shubin-Vonsovsky model taking into account not only the Hubbard
repulsion, but also the intersite Coulomb interactions. It is
shown that the account of the Kohn-Luttinger renormalizations up
to the second order of perturbation theory inclusively and the
allowance for the Coulomb repulsion between electrons located at
the neighboring and the next-nearest neighboring carbon atoms
determine to a considerable extent the competition between the
$f$-, $p+ip$-, and $d + id$-wave superconducting phases. They also
lead to a significant increase in the absolute values of the
effective interaction and, hence, to the higher superconducting
transition temperatures for the idealized graphene monolayer.

Note that for the $p$-, $d$-, $f$-wave, as well as for the
$s$-wave pairing with nodal points in 2D
($\Delta_s(\phi)\sim\cos\,6n\phi,\,\Delta_{s_{ext}}(\phi)\sim\sin\,6n\phi,\,n\geq1$),
the Anderson theorem for non-magnetic impurities is violated and
anomalous superconductivity is totally suppressed for $\gamma\sim
T_c^{clean}$, where $\gamma$ is an electron damping due to the
impurity scattering ($\gamma=1/(2\tau)=\pi
n_{imp}|V_{el-imp}(0)|^2\rho_{2D}(\mu)$ in the Born
approximation~\cite{Posazhennikova96}).

\begin{acknowledgements}
The authors thank V.V. Val'kov for stimulated discussions. The
work is supported by the Russian Foundation for Basic Research
(nos. 14-02-00058 and 14-02-31237). One of the authors
(M.\,Yu.\,K.) thanks support from the Basic Research Program of
the National Research University Higher School of Economics. The
work of another one (M.\,M.\,K.) is supported by the scholarship
SP-1361.2015.1 of the President of Russia and the Dynasty
foundation.
\end{acknowledgements}

\end{document}